\documentclass[paper,twocolumn,showpacs,superscriptaddress,apl]{revtex4}%
\usepackage{amsfonts}
\usepackage{amsmath}
\usepackage{amssymb}
\usepackage{graphicx}%
\usepackage{comment}

\begin{document}
\title{Gate-tunable nearly total terahertz absorption in graphene with resonant metal back reflector}

\author{Jiang-Tao Liu}
\email{jtliu@semi.ac.cn}
\address{Nanoscale Science and Technology  Laboratory,  Institate for Advanced Study,  Nanchang University, Nanchang 330031, China}
\address{Department of Physics, Nanchang University, Nanchang 330031, China}
\author{Nian-Hua Liu}
\address{Nanoscale Science and Technology  Laboratory,  Institate for Advanced Study,  Nanchang University, Nanchang 330031, China}
\address{Department of Physics, Nanchang University, Nanchang 330031, China}
\author{Li Wang}
\address{Nanoscale Science and Technology  Laboratory,  Institate for Advanced Study,  Nanchang University, Nanchang 330031, China}
\address{Department of Physics, Nanchang University, Nanchang 330031, China}
\author{Xin-Hua Deng}
\address{Department of Physics, Nanchang University, Nanchang 330031, China}

\author{Fu-Hai Su}
\email{fhsu@issp.ac.cn}
\address{Key Laboratory of Materials Physics, Institute of Solid State Physics, Chinese Academy of
Sciences, Hefei 230031, China}



\date{\today}

\begin{abstract}
The gate-tunable terahertz (THz) absorption of graphene layers with a resonant metal back reflector (RMBF) is theoretically investigated. We demonstrate that the THz absorption of graphene with RMBF can vary from nearly negligible to nearly total by tuning the external gate voltage. This peculiar nearly total THz absorption can be attributed to the Fabry-Perot cavity effect, which enhances the absorption and reduces the reflection of graphene. The absorption spectra of the graphene-RMBF structure can also be tailored in bandwidth and center frequency by changing the thickness and  dielectric constant of the spacer layer.
\end{abstract}
\maketitle

The optical properties of graphene is attracting increased attention because of the abundant potential applications within a wide spectral range from terahertz (THz) to visible frequencies \cite{1FB,2KS,3XW,4SB,ad1FW,ad1AV,7NM,8ML,9ZZ,10PH,ad2ZF,ad2NJ,12ST,13AF,14ME,15AN,ad2ZF2,16ST,17HY,18JL,ad1XZ,ad1HD,30XZ,20XZ}. As an ultra-thin two-dimensional (2D) carbon material, graphene is widely used in the transparent electrodes and optical display materials \cite{1FB,2KS,3XW,4SB}.  In recent years, THz techniques have been used to study the electric states in graphene  \cite{23HY,36ST,37PA,28IT,38JM,39IM}. Given   the ultra-high carrier mobility of graphene, it also has applications in THz optoelectronics such as transformation optics \cite{ad1AV}, tunable THz modulators \cite{24LJ,25LR,26BS,29BS,19SH},  room-temperature THz detectors \cite{27LV}, THz optical antennas \cite{22PL},    etc.  These graphene-based THz devices have important applications, such as in medical diagnostics, molecular biology, and homeland security.

To promote the applications of graphene within the THz  frequency range, the interaction between graphene and THz waves should be enhanced. In the recent two years, various graphene plasmonics with different microstructures have been proposed to enhance the absorption of graphene \cite{12ST,13AF,14ME,ad2ZF2,15AN,16ST,30XZ,17HY,18JL,ad1XZ}. In particular, nearly complete absorption can be achieved in periodically patterned graphene or microcavity \cite{12ST,13AF}. The concept of perfect absorbers has initiated a new research area and has important applications in optoelectronics \cite{12ST,addNIL,addNL,addCH}. However, fabricating periodically patterned graphene or placing it in an optical microcavity under current  technological conditions remains difficult.

Recently, Liu et al. proposed that the optical absorption of graphene layers on the top of a one-dimensional photonic crystal (1DPC) can be significantly enhanced within the visible spectral range because of photon localization \cite{32JT}. In a similar manner, the absorption of graphene can also be increased
within the THz spectra range \cite{33NM}. The proposed 1DPC structures can be implemented using existing technologies. However, the photonic band gap (PBG) of 1DPC limits the spectrum bandwidth for the absorption enhancement of graphene. In fact, highly conducting metal films such as aluminum, silver, and gold can effectively reflect the electromagnetic wave within a wide spectral range from the middle-infrared region to the microwave region the same as a 1DPC \cite{31NL}. Thus, the metal film can replace the 1DPC to enhance the THz absorption of graphene within a wide spectrum region.  Apart from performing as the back reflector, the metal film can also act as the metal gate electrode that can be used to modulate graphene absorption.

In this Letter, the THz absorption of  graphene layer prepared on top of SiO$_{2}$/p-Si substrate with a resonant metal back reflector (RMBF)  is theoretically investigated. We find that the absorption of graphene with an RMBF can be enhanced by about 3.3 times because of Fabry-Perot interference. The full width at half maximum of the absorption spectrum (FWHM) of graphene with an RMBF is much larger than that of graphene on top of a 1DPC. By tuning the applied gate voltage, the THz absorption of graphene layer with an RMBF can vary from nearly total transparency to total absorption regardless of the incident angle if  this angle is not too large. Our proposal is very easy to implement using the existing technology and has potential important applications in both THz and graphene studies.

The details of the structure are shown in the inset of Fig. 1. The spacer layers consist of 300 nm SiO$_{2}$ layers and a 21.9 $\mu m$ lightly doped p-type silicon (p-Si) layer with resistivity greater than 100 $\Omega$ cm, unless otherwise specified. The graphene layer is prepared on top of the SiO$_{2}$ layer, and an 86 nm silver film is placed at the bottom of the p-Si layer as the back reflector and metal gate electrode \cite{addsm}. The refractive index of the p-Si (SiO$_{2}$) layer is 3.418 (2.1) and negligible in THz absorption \cite{32DG}.  The conductivity of Ag film is $\sigma_{Ag}=46(\mu\Omega)^{-1}$, and the complex refractive index is $n_{Ag}=(1+i)\sqrt{\sigma_{Ag}/(4\pi\varepsilon_{0}f)}$, where $\varepsilon_{0}$ is the vacuum dielectric constant, and $f$  the THz wave frequency \cite{31NL}. Within the  THz frequency range, the conductivity of graphene can be expressed as \cite{33NM}
\begin{equation}
\sigma_{g}=\frac{e^{2}}{\pi \hbar }\frac{|\epsilon _{F}|}{\hbar \Gamma -i\hbar \omega},
\end{equation}
where $\hbar$ is the reduced Planck constant,  $\Gamma$ is the relaxation rate, $\epsilon _{F}$ is the Fermi
level position with respect to the Dirac point, and $\omega$ is the angular frequency
 of the incident THz radiation.

To model the THz absorption of  graphene in this structure, the transfer matrix method is used \cite{32JT,34MB,35KC}. The electric field of the TE mode and the magnetic field of the TM mode of THz waves in the \emph{l}th layer is given by $E_{l}(y,z)=(A_{l}e^{ik_{z}z}+B_{l}e^{-ik_{z}z})e^{-ik_{y}y}e_{x}$ and $H_{l}(y,z)=(A_{l}e^{ik_{z}z}+B_{l}e^{-ik_{z}z})e^{-ik_{y}y}e_{x}$, respectively. Where  the TM (TE) mode is defined as the component of the magnetic (electric) field parallel to the graphene layers. The relation of the electromagnetic field in the \emph{l}th layer to the incident electromagnetic wave is
\begin{equation}
\left(
\begin{array}{c}
A_{l} \\
B_{l}%
\end{array}%
\right) =\left(
\begin{array}{cc}
T_{11} & T_{12} \\
T_{21} & T_{22}%
\end{array}%
\right) \left(
\begin{array}{c}
A_{0} \\
B_{0}%
\end{array}%
\right). \label{eqn1}
\end{equation}
Thus, we can obtain  the absorbance of graphene $\mathcal{A}_{o}$  using the Poynting vector \cite{32JT}
\begin{equation}
\mathcal{A}_{o}=(\mathcal{S}_{0i}+\mathcal{S}_{2i}-\mathcal{S}_{0o}-\mathcal{S}_{2o})/\mathcal{S}_{0i},
\end{equation}
where $\mathcal{S}_{0i}$ and $\mathcal{S}_{0o}$ ($\mathcal{S}_{2i}$ and $\mathcal{S}_{2o}$) are the incident  and  outgoing Poynting vectors in air (in the nearest spacer layer), respectively. Here, $\mathcal{S}_{0i}=\beta_{0}A_{0}^{2}\cos\theta$, $\mathcal{S}_{0o}=\beta_{0}B_{0}^{2}\cos\theta$, $\mathcal{S}_{2i}=\beta_{1}B_{2}^{2}\cos\theta'$, and $\mathcal{S}_{2o}=\beta_{1}A_{2}^{2}\cos\theta'$. For the TE mode, $\beta_{0}=\sqrt{\varepsilon_{0}/\mu_{0}}$, $\beta_{1}=\sqrt{\varepsilon_{s}/\mu_{0}}$; for the TM mode, $\beta_{0}=\sqrt{\mu_{0}/\varepsilon_{0}}$, $\beta_{1}=\sqrt{\mu_{0}/\varepsilon_{s}}$, where $\varepsilon_{s}$ is the dielectric constant of the spacer layer, and $\theta'$ is the propagation angle of light in the spacer layer.

\begin{figure}[t]
\includegraphics[width=0.9\columnwidth,clip]{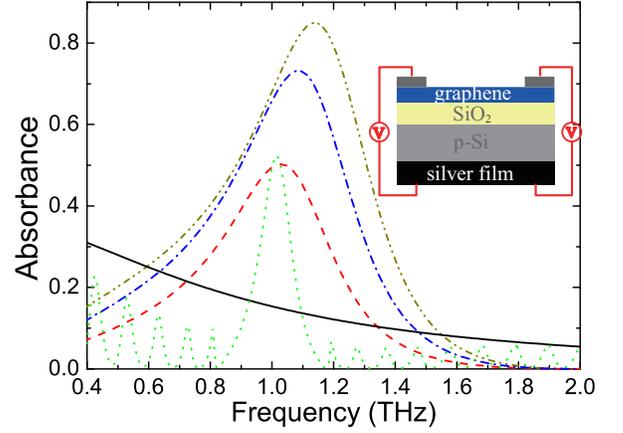}
\caption{(color online) The absorbance of graphene as a function of the frequency
for different structures: suspended   graphene monolayer
(black solid line), graphene monolayer with a 1DPC (green dotted line),
graphene monolayer with an RMBF (red dashed line),
two graphene monolayers with an RMBF (blue dash-dotted  line), and three graphene monolayers (dark yellow dash-dot-dotted line).
 The inset shows the schematic of the graphene layer prepared on
top of SiO$_{2}$/p-Si spacer layers with an RMBF.}%
\label{fig1}%
\end{figure}

Figure 1 shows the THz absorption of graphene layers under normal incidence in different structures  as a function of frequency. In the calculations, $\hbar \Gamma =2.5$ meV and $\varepsilon_{F}=0.06$ eV are used.  The maximum THz absorbance of the graphene monolayer with an RMBF is about 0.5 (red dashed line in Fig. 1). By contrast, for the same frequency ($f\approx 1$ THz), the absorbance of the suspended graphene monolayer is about 0.15 (black solid line in Fig. 1). Thus, the absorption of graphene monolayer with an RMBF can be enhanced by about 3.3 times. For three graphene monolayers, a maximum THz absorbance of 0.85 can be achieved (dash-dot-dotted line in Fig. 1). Similar to graphene with a 1DPC, the graphene layer and the metal film act as the mirrors of the Fabry-Perot Cavity. The THz wave propagates back and forth between these two mirrors, which  leads to photon localization and enhances the absorption of graphene \cite{32JT,35KC}.

In Fig. 1, we also show the THz absorption of graphene layers prepared on top of 7.5 period alternating Si and SiO$_{2}$ layers, i.e., 1DPC. The maximum THz absorbance of the graphene monolayer with a 1DPC is about 0.52 (green dotted line in Fig. 1), which is slightly larger than that of the graphene monolayer with an RMBF. However, the FWHM of the absorption spectrum of graphene monolayer with a 1DPC is only about 0.11 THz  limited to the PBG width in 1DPC. By contrast, the bandwidth of  absorption in the graphene-RMBF structure can  reach 0.45 THz because the metal film can perfectly reflect the electromagnetic wave  within the wide spectral range. More importantly,  graphene-RMBF structure  is much easier to realize than other proposed structures such as 1DPC, periodically patterned graphene, and microcavity. THz spectroscopy is also used to detect electron states and ultrafast dynamics of Dirac fermions in graphene \cite{23HY,36ST,37PA,38JM,39IM,28IT}.  The enhanced THz absorption with an RMBF can promote these studies and may help  observe the THz-induced nonlinear dynamics of Dirac fermions in graphene \cite{39IM,40FB}.

The THz absorption of graphene layers can be tuned  by varying the gate voltage. The Fermi energy $|\varepsilon_{F}|$ of graphene can be  continuously tuned by varying the gate-voltage similar to a field-effect transistor. As shown in Eq. (1), the conductivity of graphene is expected to increase as the Fermi energy $|\varepsilon_{F}|$ is increased, which  enhances the   intraband THz absorption of graphene.  By using a simple capacitor model, the Fermi energy $|\varepsilon_{F}|$ of graphene in our proposed structure can be written as\cite{ad1FW,25LR}
\begin{equation}
|\varepsilon_{F}|=\hbar v_{F}\sqrt{\pi|\alpha_{c}(V_{g}-V_{E0})|},
\end{equation}
where $v_{F}=1\times 10^{6}$ m/s is the Fermi velocity,  $\alpha_{c}\approx 7\times 10^{10}$ cm$^{-2}$V$^{-1}$,  $V_{E0}=\varepsilon_{F0}^{2}/(\hbar^{2} v_{F}^{2}\pi\alpha_{c})$,  and $\varepsilon_{F0}$ is the  Fermi energy of graphene with zero gate-voltage.

\begin{figure}[t]
\includegraphics[width=0.85\columnwidth,clip]{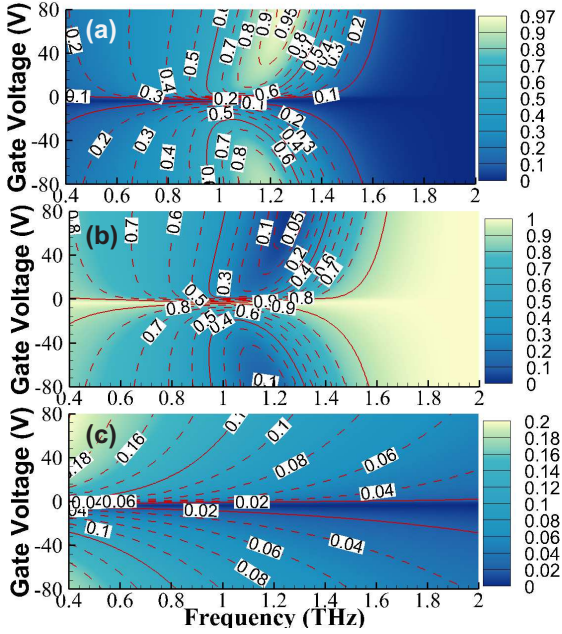}
\caption{(color online). Contour plots of the (a)  absorbance and (b) reflectance of  graphene monolayer
with an RMBF as a function of the light frequency and  back gate voltage.
(c) Contour plots of  the absorbance  of graphene monolayer without an RMBF as a function of the light frequency
 and back gate voltage.}%
\label{fig2}%
\end{figure}

The absorbance and reflectance of  graphene monolayer with an RMBF as a function of the light frequency and  back gate voltage for $\varepsilon_{F0}=0.06$ eV are shown in Fig. 2(a) and 2(b), respectively. The absorbance of graphene with an RMBF can vary from nearly zero to nearly 100\% by  tuning the gate voltage. For instance, when $f=1.26$ THz, the absorbance (reflectance) of graphene with an RMBF for $V_{g}=3.8$ V, $V_{g}=45$ V, and $V_{g}=80$ V is about 0.03 (0.96), 0.9 (0.08), and 0.97 (0.02), respectively. Thus, the absorbance and reflectance are very sensitive  to the  gate-voltage.  By contrast, the absorbance of the graphene monolayer without an RMBF as a function of optical frequency and the back gate voltage is shown in Fig. 2(c).  To remove the Fabry-Perot cavity  effect, the thickness of the p-Si layer is set to semi-infinite. The maximum absorbance of graphene without an RMBF is only about 0.2, which is even smaller than that of suspended graphene monolayer (black solid line in Fig. 1). Thus, the traditional substrate material reduces the absorption of graphene. In fact, the absorption coefficient of graphene within the THz  frequency range is  large. The weak absorbance of graphene is due to the fact that most  of the incident THz wave is reflected because of the large real and imaginary parts of the conductivity of graphene for a large Fermi energy $|\varepsilon_{F}|$ \cite{addYV}. A resonant  back reflector such as 1DPC or metal can  reduce the reflection of graphene \cite{35KC, ad1BSR} and lead to relatively weak photon localization \cite{32JT}, which are the key points for achieving nearly total THz absorption in graphene.

\begin{figure}[t]
\includegraphics[width=0.98\columnwidth,clip]{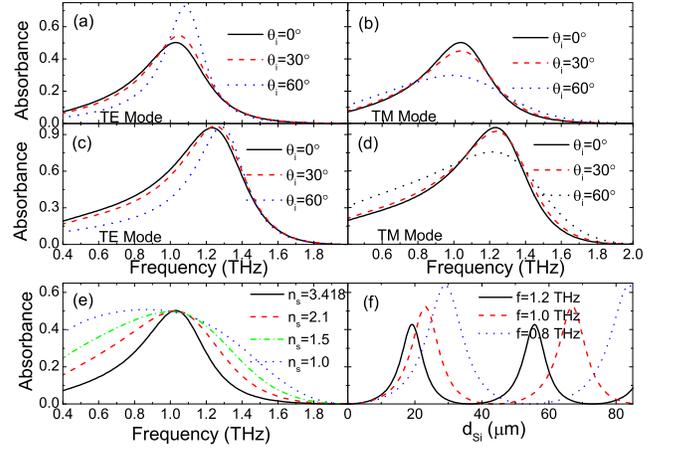}
\caption{(color online). Absorbance of graphene monolayer with an RMBF as a
function of the light frequency for different incident
angles: (a, b) $\varepsilon_{F}=60$ meV , and (c, d) $\varepsilon_{F}=300$ meV.Absorbance of graphene monolayer with an RMBF as a
function of (e) the light frequency   with  different spacer layers  and (f) the thickness of p-Si spacer layer.}
\label{fig3}%
\end{figure}

The absorption spectra of the graphene-RMBF structure in the case of oblique incidence of a THz beam are shown in Fig. 3. The resonance condition of a Fabry-Perot cavity can be described as $2L_{o}k \cos\theta'=2m\pi$, where $m$
is an integer, $L_{o}$ is the optical path of the p-Si and SiO$_{2}$ spacer layers,
 $k$ is the wave vector of the THz wave, and $\theta'$ is the propagation
angle of the THz wave in the spacer layer. According to  Snell's law,  $\theta'$ is small even with a large incident angle $\theta_{i}$ because of the large  refractive index of the p-Si layer  in the proposed structure.  Different from the 1DPC, the reflection of the metal films within the THz range is nearly invariable with increased incident angle. Thus, the   THz absorption of graphene with an RMBF is less affected by the incident angle of the THz beams if the incident angle is not too large. For instance, for $\varepsilon_{F}=300$ meV,  the absorbance of graphene monolayer  for $\theta_{i}=0^{\circ}$ and  $\theta_{i}=30^{\circ}$ for TE (TM) mode is about 0.956 (0.956) and 0.972 (0.926). Even for incident angle $\theta_{i}=60^{\circ}$, the absorbance of graphene monolayer  for the TE (TM) mode is about 0.963 (0.757).  Thus, our proposed graphene-RMBF structure can be used in pantoscopic and imageable THz detectors and modulators.

We now consider the adjustability of the THz absorption of graphene by varying the dielectric constant or optical path of the spacer layers. The reflection of the graphene layer is smaller with a lower dielectric constant substrate, and the THz absorption of graphene is enhanced beyond the resonance cavity frequency.  The FWHM of the THz absorption spectra increases with decreased dielectric constant of the spacer layers [see Fig. 3(e)]. For the normal incidence case, the resonance condition of a Fabry-Perot cavity is $2L_{o}k=2m\pi$, which indicates that the center frequency of the absorption spectra can be easily tuned by varying the spacer layers thickness. As shown in Fig. 3(f), the center frequency of the THz absorption peak increases with  decreased spacer layers thickness. Given that the absorption coefficient of graphene is larger for lower frequency THz waves according to Eq. (1), the maximum absorption of graphene increases with increased spacer layers thickness. Thus, the fabrication of graphene-RMBF structure  with a thickness-tunable  air spacer layer (e.g.,  gate controlled suspended graphene \cite{41WB1,42WB2}) is also highly desired.

In summary, the gate-tunable THz absorption of  graphene layers with an RMBF is investigated.  The THz absorption of graphene with an RMBF is enhanced and can be tuned from nearly zero to nearly total by varying the gate voltage. This peculiar nearly total THz absorption can be attributed to the Fabry-Perot cavity effect, which enhances the absorption of graphene and reduces the reflection of the THz beams. The maximum absorption of graphene is almost unchanged even when the incident angle varies from 0$^\circ$ to 30$^\circ$. The center frequency and  FWHM of the absorption spectra can also be tuned by varying the thickness and  dielectric constant of the spacer layers. Using existing technology, the proposed structure is very easy to fabricate not only in laboratory scale but also in industrial scale. Our findings have important implications in the development of THz photonic devices such as THz detectors and modulators, as well as in studies on the ultrafast dynamics of Dirac fermions in graphene.

This work was supported by the National Key Basic Research Program of China (2013CB934200), the NSFC Grant Nos. 11104232, 11264030, and 11364033, the NSF from the Jiangxi Province Nos.  20122BAB212003, and Science and Technology Project of Education Department of Jiangxi Province No. GJJ13005.


\begin{thebibliography}{99}
\bibitem {addsm}Changes in THz absorption are negligible if the 86 nm silver films are replaced by Al, Au, or  thicker silver films.
\end{thebibliography}
\end{document}